\documentstyle[aps,epsf,twocolumn]{revtex}

\def\mev{\rm meV}
\def\nlsm{NL$\sigma$M\ }
\def\be{\begin{equation}}
\def\ee{\end{equation}}
\def\bea{\begin{eqnarray}}
\def\eea{\end{eqnarray}}

\def\t12h{\frac{\theta_{12}}{2}}

\def\eps{\varepsilon}

\def\r#1{(\ref{#1})}
\def\nn{\nonumber\\}

\def\NPB#1#2#3{{ Nucl. Phys.} {B{\bf #1}} #2 (#3)}
\def\PRD#1#2#3{{ Phys. Rev.} {D{\bf #1}} #2 (#3)}

\def\d{\Delta}

\begin{document}

\preprint{OUTP-99-??S}

\title{\bf On the 3-particle scattering continuum in quasi
one dimensional integer spin Heisenberg magnets}
\author {Fabian H.L. Essler}
\address{Department of Physics, Theoretical Physics,
        Oxford University\\ 1 Keble Road, Oxford OX1 3NP, United Kingdom}

\maketitle
\begin{abstract}
We analyse the three-particle scattering continuum in quasi
one dimensional integer spin Heisenberg antiferromagnets within a
low-energy effective field theory framework. We exactly determine the
zero temperature dynamical structure factor in the O(3) nonlinear sigma
model and in Tsvelik's Majorana fermion theory. We study the effects of
interchain coupling in a Random Phase Approximation. We discuss the
application of our results to recent neutron-scattering experiments on
the spin-1 Haldane-gap material ${\rm CsNiCl_3}$. 
\end{abstract}
\date{today}
\begin{narrowtext}

\section{Introduction}

In recent inelastic neutron scattering experiments \cite{mr} on the
quasi one dimensional spin-1 Heisenberg magnet ${\rm CsNiCl_3}$
\cite{csnicl3}, the existence of incoherent multiparticle scattering
continua in the one dimensional phase was investigated. If was found,
that close to the antiferromagnetic wave vector (along the chain
direction), there is significant spectral weight above the coherent
magnon peak. 

Motivated by these experimental results we determine the dynamical
structure factor for weakly coupled integer spin Heisenberg chains in
a low-energy effective field theory framework. In particular, we
calculate the ratio of spectral weights of multiparticle scattering
continua to the coherent magnon peak. 

An appropriate model Hamiltonian for ${\rm CsNiCl_3}$ is \cite{csnicl3}
\be
H=J\sum_{<ij>} \vec{S}_{i}\cdot\vec{S}_{j}
+J^\prime\sum_{<ij>^\prime}\vec{S}_{i}\cdot\vec{S}_{j}
+D\sum_{n} (S^z_n)^2\ ,
\label{hamil}
\ee
where the first and second sums are over nearest neighbour spins along
and between the chains respectively. The exchange constants are
estimated to be 
$J\approx 2.8\mev$, $J^\prime\approx 0.045\mev$
and the single-ion anisotropy is estimated to be $D\approx -0.004\mev$
\cite{csnicl3}. 
Above the ordering temperature of about ${\rm 4.84 K}$, ${\rm
CsNiCl_3}$ is considered to be a good realisation of a one dimensional
Haldane-gap system, although there is a sizeable dispersion
perpendicular to the chain direction due to the interchain
coupling. As $D$ is very small, we neglect it from now on.

As a first approximation we can neglect the effects due to $J^\prime$,
so that we arrive at the purely one dimensional Heisenberg
Hamiltonian 
\be
H_{\rm 1D}=J\sum_{n} \vec{S}_{n}\cdot\vec{S}_{n+1}\ .
\label{h1d}
\ee
The dynamical susceptibility for \r{h1d} has been calculated
numerically for a chain of 20 sites by Takahashi \cite{taka}. He found,
that about three percent of the total intensity above the
antiferromagnetic wave vector are due to incoherent multiparticle
scattering continua. A recent dynamical density matrix renormalisation
group calculation an large systems of up to $320$ sites quotes a
result of about $2.5$ percent \cite{kw} for this quantity.
The purpose of the present work is to determine the incoherent
contribution to the structure factor by analytical means in the
thermodynamic limit. At present, the best way to do this 
is by using a low-energy effective field theory description of \r{h1d}
and taking the interchain coupling in \r{hamil} into account perturbatively.
The dynamical magnetic susceptibility for a spin chain is given by
\bea
\chi^{\alpha\beta}(\omega,q)&=& \frac{-i}{a_0}\int_{-\infty}^\infty\!\!
dx\int_0^\infty\!\! dt\ e^{i\omega t-iqx}\nn
&&\times\quad
\langle\left[S^\alpha(t,x),S^\beta(0,0)\right]\rangle .
\eea
Note that we use units in which $\hbar=1$.
The dynamical structure factor is obtained as (see e.g. \cite{dms})
\be
S^{\alpha\beta}(\omega,q)=-\frac{1}{\pi}{\rm Im}\
\chi^{\alpha\beta}(\omega,q)\ .
\label{dsf}
\ee
Due to the spin-rotational symmetry of \r{h1d}, we can restrict our
attention to the case $\alpha=\beta=z$. We denote the corresponding
susceptibility and structure factor by $\chi(\omega,q)$ and
$S(\omega,q)$, respectively. Below we will calculate the structure
factor in two different low-energy effective field theories. We
concentrate on the region $q\approx\pi$, which is of experimental
relevance. It is known that in this region there exists a coherent one
magnon excitation with a gap $\Delta$ and incoherent three, five, seven etc
magnon scattering continua. We constrain our analysis to the case of
three magnons, which gives the dominant contribution.
We also investigate the effects of the coupling between chains, but do
not take into account the single-ion anisotropy. We note that
the analysis of section \ref{sec:maj} can be easily extended to the
case $D\neq 0$.

The outline of this paper is as follows. In section \ref{sec:nlsm} we
determine the dynamical structure factor of the one dimensional model
\r{h1d} for general integer spin $S$ in the framework of the O(3)
nonlinear $\sigma$-model (NL$\sigma$M) description. In section
\ref{sec:rpa} we study the effects of the coupling between chains in
\r{hamil}, using the \nlsm results as an input. In section
\ref{sec:maj} we determine the structure factor within a second
low-energy effective field theory description, which holds for $S=1$
in the presence of a (strong) biquadratic exchange interaction \cite{biquad}
between spins in addition to \r{hamil}. Although the structure of the
excitation spectrum as a function of wave number along the chain
direction is essentially the same as in the \nlsm, the structure
factor turns out to be rather different. In section
\ref{sec:temperature} we discuss temperature effects and in section
\ref{sec:summary} we summarise and discuss our results.

\section{O(3) Nonlinear Sigma Model}
\label{sec:nlsm}
At energies much smaller than the exchange $J$, the lattice model
\r{h1d} (for integer spin $S$) can be approximated by a field theory, the
O(3) \nlsm (see e.g. \cite{affleck88,fradkin}). 
The Lagrangian density of the \nlsm is given by
\be
{\cal L}=\frac{1}{2g}\left[\frac{1}{v}\left(\frac{\partial
m^a}{\partial t}\right)^2-
{v}\left(\frac{\partial
m^a}{\partial x}\right)^2\right]\ ,\quad \vec{m}\cdot\vec{m}=1\ ,
\label{lagr}
\ee
where $\vec{m}$ is a three-component vector and $v$ is the magnon velocity.
The relation between lattice and field theory variables is given by
\cite{affleck88,fradkin}
\be
\vec{S}(x)\approx S (-1)^{x/a_0} \vec{m}(x) +\frac{1}{vg}
\vec{m}(x)\times\frac{\partial\vec{m}(x)}{\partial t}\ ,
\label{decomp}
\ee
where $a_0$ is the lattice spacing.
In order for the NL$\sigma$M to give an accurate description of the
low-energy physics of the spin chain, the spin $S$ is supposed to be
large. However, previous investigations \cite{affleck} indicate that
even the case $S=1$ is described rather well by the NL$\sigma$M. 
Using \r{decomp}, it is possible to determine dynamical correlation
functions within the framework of the \nlsm. The $q\approx 0$ behaviour of
the dynamical structure factor was determined in this way by Affleck
and Weston some time ago \cite{aw}. For $q\approx 0$ the \nlsm
predicts an incoherent two magnon scattering continuum. As the
scattering intensity is proportional to $q^2$, this continuum is very
difficult to observe experimentally.

Here we are interested in wave-number transfers close to the
antiferromagnetic point $q\approx\frac{\pi}{a_0}$. Using \r{decomp},
we see that the dynamical susceptibility is given by
\bea
\chi(\omega,q)&=& \frac{-iS^2}{a_0}\int_{-\infty}^\infty\!\!\!
dx\int_0^\infty\!\! dt e^{i\omega t-iqx}\nn
&&\qquad \times \langle\left[m_3(t,x),m_3(0,0)\right]\rangle\ .
\label{chi}
\eea
The two-point function $\langle m_a(t,x)\ m_a(0,0)\rangle$
has been calculated using the formfactor approach \cite{ff} in
\cite{ks,bn}. 
Let us briefly review some relevant formulas. The exact spectrum of
the O(3) \nlsm consists of three massive magnons $A_a$, $a=1,2,3$,
that form the vector representation of O(3). We denote the magnon mass
gap by $M$. Due to factorisability of the exact scattering matrix
\cite{zz}, multi magnon scattering states form a basis of the Hilbert
space. Let us introduce some notations. We parametrise energy and
momentum of one magnon states in terms of a rapidity variable $\theta$
\be
E_1=Mv^2\cosh\theta\ ,\quad P_1=Mv\sinh\theta\ .
\ee
A scattering state of $N$ magnons with rapidities $\{\theta_j\}$ and O(3)
indices $\{\eps_j\}$ ($\eps_j=1,2,3$) is denoted by 
\be
|\theta_1,\theta_2,\ldots,\theta_N\rangle_{\eps_1,\eps_2,\ldots \eps_N}\
.
\ee
Its energy and momentum are
\be
E_N=Mv^2\sum_{j=1}^N\cosh\theta_j\ ,\quad P_N=Mv\sum_{j=1}^N\sinh\theta_j\ .
\ee
The resolution of the identity is given by
\bea
1&=&\sum_{n=0}^\infty\frac{1}{n!}\
\sum_{\{\eps_j\}}
\int_{-\infty}^\infty \prod_{j=1}^n\frac{d\theta_j}{4\pi}\nn
&&\times\ |\theta_1,\ldots,\theta_n\rangle_{\eps_1,\ldots \eps_n}\
^{\eps_1,\ldots \eps_n}\langle\theta_1,\ldots,\theta_n|\ .
\eea
The two point function of some operator ${\cal O}$ can now be
expressed in the spectral representation as
\bea
&&\langle{\cal O}^\dagger(t,x){\cal O}(0,0)\rangle=
\!\sum_{n=0}^\infty\frac{1}{n!}
\!\int_{-\infty}^\infty
\prod_{j=1}^n\frac{d\theta_j}{4\pi}
|{\cal F}_{\cal O}^{(n)}(\theta_1,\ldots,\theta_n)|^2\nn
&&\times\exp\left(-it\sum_{k=1}^nv^2M\cosh\theta_k
+ix\sum_{k=1}^nvM\sinh\theta_k\right).
\label{2pt}
\eea
Here 
\bea
&&|{\cal F}_{\cal O}^{(n)}(\theta_1,\ldots,\theta_n)|^2=\nn
&&\quad\sum_{\{\eps_j\}}
|\langle 0|{\cal
O}^\dagger(0,0)|\theta_1,\ldots,\theta_n\rangle_{\eps_1\ldots\eps_n}|^2. 
\label{ME}
\eea
The matrix elements for several operators ${\cal O}$ in \r{ME}
have been determined in \cite{ks,bn}. In order to calculate the
structure factor, we are interested in matrix elements of (any of the
components of) the fundamental field of the \nlsm, e.g. ${\cal O}=m_3$.
The summed absolute values of the first few matrix elements for this
case are 
\bea
&&|{\cal F}^{(2k)}(\theta)|^2=0\ ,\quad k=1,2,\ldots,\nn
&&|{\cal F}^{(1)}(\theta)|^2={\bf Z}\ ,\nn
&&|{\cal F}^{(3)}(\theta_1,\theta_2,\theta_3)|^2={\bf Z}
\pi^6\left[12\pi^2+2\sum_{j<i}(\theta_j-\theta_i)^2\right]\nn
&&\prod_{k<l} \frac{(\theta_k-\theta_l)^2+\pi^2}{(\theta_k-\theta_l)^2
([\theta_k-\theta_l]^2+4\pi^2)}\left(\tanh\frac{\theta_k-\theta_l}{2}\right)^4.
\label{matrixel}
\eea
Here the overall factor $\bf Z$ is due to the field renormalisation in
the \nlsm. In order to determine the dynamical susceptibility, we now
use \r{matrixel} and \r{2pt} in \r{chi} and then carry out the $x$ and
$t$ integrations. In this way we arrive at
\bea
&&\chi(\omega,q)=\frac{{\bf Z}^\prime}{\omega^2-v^2q^2-\d^2+i\eps}\nn
&&+\frac{\pi^6{\bf Z}^\prime}{3}\int_{-\infty}^\infty\frac{dz}{4\pi}
\int_{-\infty}^\infty\frac{dy}{4\pi}\left[12\pi^2+4(3z^2+y^2)\right]\nn
&&\times \frac{f(y+z)f(y-z)f(2z)}{\omega^2-v^2q^2-\d^2-4\d^2\cosh
z(\cosh z+\cosh y)+i\eps}\nn
&&{\rm +\ contributions\ from\ 5,7,9...\ particles},
\label{chiff}
\eea
where ${\bf Z^\prime}=S^2 Z v/a_0$, $\Delta= Mv^2$ and
\be
f(z)=[\tanh(z/2)]^4\ \frac{z^2+\pi^2}{z^2[z^2+4\pi^2]}\ .
\ee
The structure factor close to the antiferromagnetic wave number is
obtained from this using \r{dsf}
\bea
&&S(\omega,q)=\frac{{\bf
Z}^\prime}{2\sqrt{v^2q^2+\d^2}}\ \delta(\omega-\sqrt{v^2q^2+\d^2}) \nn
&&+\frac{\pi^4{\bf Z}^\prime}{3}\int_{0}^{z_0}\!\!{dz}
\left[3\pi^2+3z^2+Y^2\right]f(2z)f(z+Y)f(z-Y)\nn
&&\times\frac{1}
{\sqrt{\left[(\omega^2-v^2q^2-\d^2-4\d^2\cosh^2 z)^2-16 \d^4 \cosh^2
z\right]}}\nn 
&&{\rm +\ contributions\ from\ 5,7,9...\ particles}.
\label{sf}
\eea
Here $z_0={\rm arccosh}\frac{x-1}{2}$ and $Y={\rm
arccosh}\frac{x^2-1-4\cosh^2 z}{4\cosh z}$, where
$x^2=(\omega^2-v^2q^2)/\d^2$. An important difference between \r{sf}
and \r{chiff} is the way, in which states with $3,5,7...$ magnons
contribute. It can be easily seen from the upper limit of integration,
that three magnon states contribute to the structure factor only if
$s^2:=\omega^2-v^2q^2> 9\d^2$, i.e. above the three magnon
threshold. Similarly, $2n+1$ magnon states only contribute if
$s>(2n+1)\d$. In other words, \r{sf} is exact as long
as $s<5\d$. The situation is quite different for ${\rm
Re}\chi(\omega,q)$: here multi magnon states contribute for all values
of $s$. However, their contribution is negligible at small $s$ (see
e.g. \cite{bn}).
The first contribution in \r{sf} is due to the coherent one magnon
states with dispersion $\omega=\sqrt{v^2q^2+\d^2}$, which is known to
be a good approxmation to the lattice-dispersion as long as $q$ is
sufficiently small (recall that in our notations $q$ denotes the
deviation from $\pi/a_0$). The total spectral weight cannot be calculated
within the \nlsm framework, so that one has to resort to a direct
numerical analysis of the quantum spin chain. For the spin-1 case this
yields \cite{affleck} ${\bf Z^\prime}\approx 1.28\frac{S^2v}{a_0}$,
$v\approx 2.5 Ja_0$.

The remaining integral in the three magnon contribution to \r{sf} can
be evaluated numerically. The result is shown in
Fig .~\ref{fig:3nlsm}.

\begin{figure}[ht]
\noindent
\epsfxsize=0.45\textwidth
\epsfbox{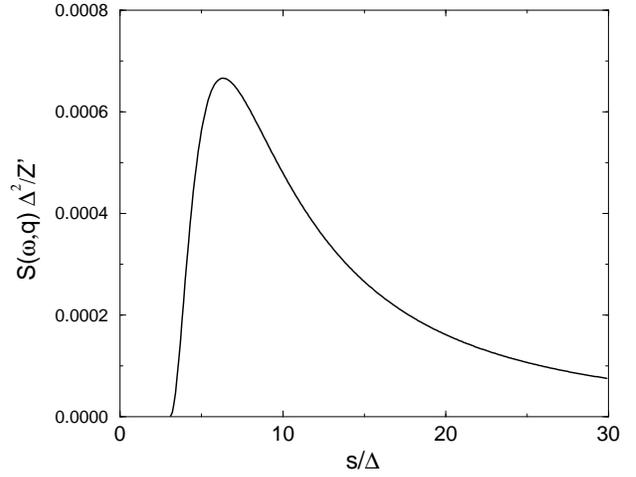}
\caption{\label{fig:3nlsm}%
\nlsm result for the three magnon contribution to the dynamical
structure factor close to the antiferromagnetic point. Here $M$ is the
magnon mass-gap, $v$ the spin velocity and $s=\sqrt{v^2q^2+\d^2}$.
} 
\end{figure}
The behaviour of the structure factor just above the three magnon
threshold can easily be determined by Taylor-expanding the integrand
in \r{sf}. The result is
\be
S(\omega,q)\approx \frac{{\bf
Z}^\prime}{\d^5}\frac{\pi^7}{4^{10}\sqrt{3}}[s-3\d]^3\ ,\quad 0<\frac{s}{\d}-3\ll 1\ .
\ee
The ratio of total spectral weights of the one magnon ($I_1$) and
three magnon ($I_3$) contributions can be calculated as
well. Numerically we find that that ratio of spectral weights at wave
number $\pi/a_0$ is roughly equal to
\be
\frac{I_3(\pi)}{I_1(\pi)}\approx 0.02\ .
\label{ratio}
\ee
We note that a sizeable fraction of the three magnon spectral weight
is located at very high energies, where the field theory does not
relate to the lattice spin model. If we restrict $\omega$ to be smaller
than twenty times the magnon gap, the ratio \r{ratio} diminishes to
about $0.013$.

The \nlsm result is somewhat at odds with the numerical results
\cite{taka,kw} on the spin 1 chain. The most likely reason is
that the \nlsm does not work as well for $S=1$ as previosuly thought.
Irrespective of the one percent difference between \nlsm and numerical
results it seems clear that the three magnon continuum of a single
spin-1 chain is extremely weak! In quasi one dimensional materials
like ${\rm CsNiCl_3}$ such a faint contribution would hardly be
measurable experimentally as it would go under in the inevitable
errors due to e.g. background subtraction. 
Another relevant quantity as far as the experiments of \cite{mr} are
concerned, is the ratio of three and one magnon spectral weights for
$q\neq 0$. We find that e.g.
\be
\frac{I_3([1\pm 0.2]\pi)}{I_1([1\pm 0.2]\pi)}\approx 0.04\ .
\label{ratio2}
\ee
The increase in \r{ratio2} as compared to \r{ratio} is mainly due to the
decrease in spectral weight of the single magnon peak
\be
\frac{I_1([1\pm 0.2]\pi)}{I_1(\pi)}\approx 0.45\ .
\ee

Let us now determine the real part of the dynamical susceptibility,
which we will need in section \ref{sec:rpa}.
Neglecting contributions of more than three magnons we
arrive at the results shown in Fig.~\ref{fig:re3nlsm}.
\begin{figure}[ht]
\noindent
\epsfxsize=0.45\textwidth
\epsfbox{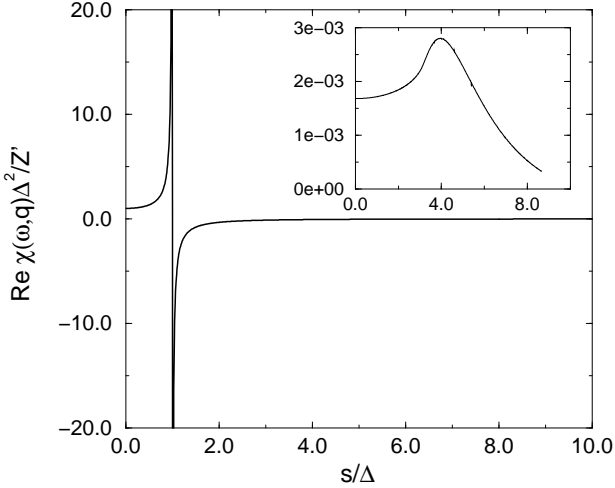}
\caption{\label{fig:re3nlsm}%
Minus the real part of the dynamical susceptibility calculated in the
framework of the \nlsm. Contributions of states with more than three
magnons are neglected. The inset shows the contribution due to three
magnon states.
} 
\end{figure}
We note that the only singularity in the real part is at $s=\d$.

\section{Coupling between the chains}
\label{sec:rpa}
Materials like ${\rm CsNiCl_3}$ are only approximately
one dimensional. There always is a small coupling between the
individual chains, which we will now take into account within the
framework of a Random-Phase Approximation (RPA). Denoting the Fourier
transform of the interchain coupling matrix elements by $J(\vec{k})$,
the three-dimensional dynamical susceptibility in RPA is simply given by
\cite{RPA}
\be
\chi(\omega,q,\vec{k})=\chi(\omega,q)[1-J(\vec{k})\chi(\omega,q)]^{-1}
\ .  
\label{chi3d}
\ee
Note that this generally is a matrix equation. In our case \r{chi3d}
reduces to a scalar equation due to SU(2) symmetry of the full,
three-dimensional Hamiltonian. We now can insert the results
\r{chiff} and \r{sf} for the dynamical susceptibility of a single
chain into \r{chi3d} and obtain in this way the dynamical structure
factor of weakly coupled chains for $q\approx \frac{\pi}{a_0}$
\be
S(\omega,q,\vec{k})=\frac{S(\omega,q)}
{|1-J(\vec{k})\chi(\omega,q)|^2}\ .
\label{sf3d}
\ee
Note that \r{sf3d} involves both the real and imaginary parts of the
one dimensional dynamical susceptibility. The magnon dispersion close
to the antiferromagnetic point along the chains is easily extracted
from the poles of \r{sf3d}
\be
\omega^2=v^2q^2+\d^2+{\bf Z}^\prime J(\vec{k})\ .
\label{magnondisp}
\ee
The coherent one magnon contribution to the structure factor is found
to be
\bea
&&S(\omega,q,\vec{k})\big|_{\rm 1\ magnon}=\frac{{\bf
Z}^\prime}{2\sqrt{v^2q^2+\d^2+{\bf Z}^\prime J(\vec{k})}}\nn
&&\qquad \times\ \delta\left(\omega-
\sqrt{v^2q^2+\d^2+{\bf Z}^\prime J(\vec{k})}\right) .
\label{1magsw}
\eea
We see that the total spectral weight due to one magnon processes
depends on the transverse wave-number transfer $\vec{k}$. 

From now on we consider the particular case of wave-number transfer
$\vec{k}=(\eta,\eta,0)$ along the (1,1,0) direction in ${\rm 
CsNiCl_3}$, which is of direct experimental relevance
\cite{mr,csnicl3}. The $\eta$ dependence of $J(\vec{k})$ is of the form
\be
J(\vec{k})=2J^\prime\left(\cos 4\pi\eta +2\cos 2\pi\eta\right)\ .
\label{jk}
\ee
We note that at the special point $\vec{k}_0=(0.19,0.19,0)$ we have
$J(\vec{k})=0$, so that (in our approximation) we are dealing with an
ensemble of uncoupled chains.
By fitting \r{magnondisp} with \r{jk} to the experimentally observed
magnon dispersion along the $(1,1,0)$ direction \cite{csnicl3}, we
obtain
\be
\d\approx 1.32 \mev\ ,\qquad
{\bf Z}^\prime J^\prime\approx 0.475 \mev^2\ .
\ee
The resulting value of $J^\prime\approx 0.05 \mev$ is by construction
close to \cite{csnicl3}.
The variation of the total spectral weight $I_1(\pi,\eta)$ of the one
magnon peak with $\eta$ \r{1magsw} (with fixed wave number $\pi/a_0$ along
the chain) is shown in Fig.\ref{fig:1mag}. 
\begin{figure}[ht]
\noindent
\epsfxsize=0.45\textwidth
\epsfbox{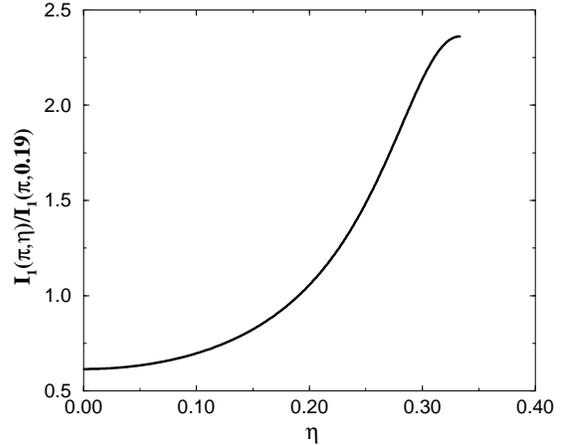}
\caption{\label{fig:1mag}%
Ratio of the total spectral weights of the one magnon peak $I_1(\pi,\eta)$
for wave-number transfer $\eta$ along the $(1,1,0)$ direction to the
``one dimensional'' result $I_1(\pi,0.19)$.
} 
\end{figure}

The three magnon contribution to the structure factor is of the form
\bea
&&S(\omega,q,\vec{k})\big|_{\rm 3\ magnons}
=\frac{S(\omega,q)\big|_{\rm 3\ magnons}}
{|1-J(\vec{k})\chi(\omega,q)|^2}\ .
\eea
Here we used the fact that the real part of the one dimensional
dynamical susceptibility did not exhibit any singularities. A result of
this is that the three magnon continuum for fixed $\eta$ starts at three
times the magnon gap only at the special point $\eta=0.19$.

From the analysis in section \ref{sec:nlsm} we know that the dominant
contribution to $\chi(\omega,q)$ for $\omega >3\d$ is due to the real
part of the one magnon contribution, which gives us an easy way to
estimate the effects of the interchain coupling on the three magnon
continuum
\bea
&&S(\omega,\pi/a_0,\vec{k})\big|_{\rm 3\ magnons}
\approx{S(\omega,\pi/a_0)\big|_{\rm 3\ magnons}}\nn
&&\quad\times\quad
\left(1-0.547\frac{\cos 4\pi\eta +2\cos 2\pi\eta}{(\omega/M)^2-1}\right)^{-2}.
\eea
In order to estimate the ratio of spectral weights of the three and
one magnon states as a function of $\eta$ we define
\be
{I_3}^\prime(\pi,\eta)=\int_0^{20M}d\omega\ S(\omega,\pi/a_0,\vec{k})\big|_{\rm 3\
magnons}\ .
\ee
Here the cutoff at twenty times the magnon gap has been chosen arbitrarily.
Note that the introduction of a cutoff is necessary to ensure the
applicability of the field-theory description.
The ratio of the three magnon spectral weight ${I_3}^\prime(\pi,\eta)$ to
the one magnon spectral weight $I_1(\pi,\eta)$ for wave-number transfer
$\eta$ along the $(1,1,0)$ is shown in Fig.\ref{fig:3mag}. 
\begin{figure}[ht]
\noindent
\epsfxsize=0.45\textwidth
\epsfbox{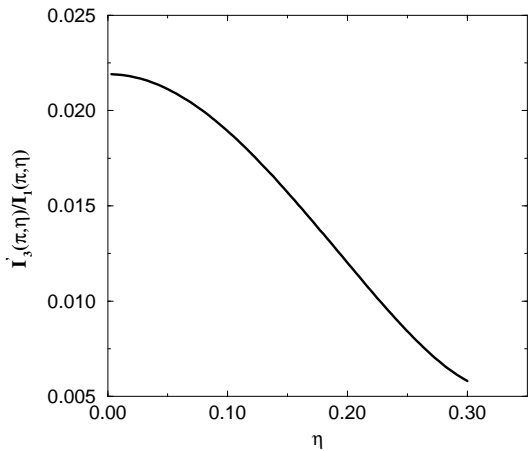}
\caption{\label{fig:3mag}%
Ratio of the total spectral weights of the three magnon continuum
${I_3}^\prime(\pi,\eta)$ for wave-number transfer $\eta$ along the $(1,1,0)$
direction to the total one magnon spectral weight ${I_1}(\pi,\eta)$.
} 
\end{figure}
We see that in all cases only a very small fraction of the total
spectral weight sits in the incoherent three magnon scattering
continuum. This fraction is at most two percent, which is much smaller
than what has been observed experimentally in ${\rm CsNiCl_3}$
\cite{mr}. We conclude that the coupling between chains cannot account
for the observed intensity in the incoherent scattering continuum at
$\pi/a_0$ as long as we use a \nlsm description for a single chain.

\section{Tsvelik's Majorana fermion theory}
\label{sec:maj}

Apart from the \nlsm there is a second low-energy effective field
theory for Haldane-gap systems, due to  A. M. Tsvelik \cite{amt}. It
applies to spin-1 models of the type 
\be
H=J\sum_{n=1}^L \vec{S}_n\cdot\vec{S}_{n+1}
-b\left(\vec{S}_n\cdot\vec{S}_{n+1}\right)^2+ D (S^z_n)^2,
\label{biq}
\ee
where $b\approx 1$. We note that one of the effects of $b\neq 0$ is to
reduce the magnitude of the Haldane gap, which rules out a very large
value of $b$ in ${\rm CsNiCl_3}$. Using the Bethe Ansatz solution of
\r{biq} at the Armenian point $b=1$, $D=0$ \cite{arm}, it is possible
to derive \cite{amt} the following low-energy effective field theory
of three interacting Majorana fermions valid for small $1-b$ and $D$
\be
{\cal L}= i \bar{\chi}_a\gamma_\mu\partial_\mu\chi_a-m_a
\bar{\chi}_a\chi_a+ g_a J^a_\mu J^a_\mu\ .
\label{maj}
\ee
Here $a=1,2,3$, $\chi_a$ are two-component Majorana (real) fermions,
$\bar{\chi}_a=\chi_a^T\gamma_0$,
$J^a_\mu=\eps^{abc}\bar{\chi}_b\gamma_u\chi_c$ are the components of
SU(2) currents and our conventions for gamma matrices are
\be
\gamma_0=\sigma_x\ ,\quad \gamma_1=i\sigma_y\ .
\ee
In the absence of a single-ion anisotropy ($D=0$), the model is SU(2)
symmetric and all Majorana masses and couplings $g_a$ are equal. We
will constrain our discussion to this case only. Following \cite{amt}
we will furthermore neglect the current-current interaction i.e. set
$g_a=0$. This is certainly justified as long as the mass gaps are not
too small and we are only interested in single-particle
properties as can be seen from a standard one-loop renormalisation
group calculation. However, it is presently not clear whether this
remains a good approximation for the calculation of the three magnon
scattering continuum. In fact, it was recently argued \cite{sach} that the
current-current interaction may become rather important for the
calculation of finite-temperature properties of spin-1/2 ladder
models, which have a very similar field-theory description
\cite{snt}. After setting $g_a=0$, \r{maj} reduces to a theory of three
noninteracting Ising models and exact information on correlation
functions is available. The staggered components of the
lattice spin operators are expressed in terms of order ($\sigma$) and
disorder ($\mu$) operators of the three Ising models as 
\be
(-1)^{x/a_0}S^x(t,x)=\sqrt{A}\sigma^1(t,x)\mu^2(t,x)\mu^3(t,x)\ ,
\ee
where the coefficient $A$ is presently not known.
The dynamical susceptibility close to the antiferromagnetic wave
number $\pi$ is thus given by
\bea
\chi(\omega,q)&=& \frac{-iA}{a_0}
\int_{-\infty}^\infty\!\!
dx\int_0^\infty\!\! dt e^{-i\omega t+iqx}\nn
&&\times\langle\left[(\sigma^1\mu^2\mu^3)(t,x),
(\sigma^1\mu^2\mu^3)(0,0)\right]\rangle\ . 
\label{chiis}
\eea
In order to determine \r{chiis} we therefore need to calculate
$(G_\mu(t,x))^2G_\sigma(t,x)$, where
\bea
G_\sigma(t,x)&=&\langle \sigma(t,x)\ \sigma(0,0)\rangle\ ,\nn
G_\mu(t,x)&=&\langle \mu(t,x)\ \mu(0,0)\rangle\ ,
\eea
are correlation functions of order and disorder operators in the
two-dimensional Ising model. These are known exactly
\cite{ising1}. For our purposes it is most convenient to work in the
Minkowski-space spectral representation, which reads \cite{ising2}
\bea
&&\langle{\cal O}^\dagger(t,x){\cal O}(0,0)\rangle=\!
\sum_{n=0}^\infty\frac{1}{n!}\!\int_{-\infty}^\infty
\prod_{j=1}^n\frac{d\theta_j}{2\pi}
|{\cal F}_{\cal O}^{(n)}(\theta_1,\ldots,\theta_n)|^2\nn
&&\times\exp\left(-it\sum_{k=1}^nv^2M\cosh\theta_k
+ix\sum_{k=1}^nvM\sinh\theta_k\right),
\label{2ptis}
\eea
where
\bea
{\cal F}_{\sigma}^{(2n+1)}(\theta_1,\ldots,\theta_{2n+1})&=&
i^\frac{2n+1}{2}\prod_{i<j=1}^{2n+1}\tanh\frac{\theta_i-\theta_j}{2}\
,\nn
{\cal F}_{\mu}^{(2n)}(\theta_1,\ldots,\theta_{2n})&=&
i^n\prod_{i<j=1}^{2n}\tanh\frac{\theta_i-\theta_j}{2}\ .
\eea
After performing the Fourier integrals we arrive at the following
representation for the dynamical structure factor
\bea
&&S(\omega,q)=\frac{vA}{\sqrt{v^2q^2+\d^2}}\
\delta(\omega-\sqrt{v^2q^2+\d^2}) \nn
&&+\frac{4vA}{\pi^2 \d^2}\int_0^{z_0}dz\frac{\tanh^2z+\frac{1}{6}\left[\tanh
z\tanh\frac{Y+z}{2}
\tanh\frac{Y-z}{2}\right]^2}{\sqrt{(x^2-1-4\cosh^2z)^2-16\cosh^2z}} 
\nn 
&&{\rm +\ contributions\ from\ 5,7,9...\ particles}\ ,
\label{sfis}
\eea
where $x^2=(\omega^2-v^2q^2)/\d^2$ and $z_0$ and $Y$ are defined above.
The one and three particle contributions in \r{sfis} give the exact
result as long as $s\leq 5\d$. For frequencies just above the
three-particle threshold the integral in \r{sfis} can be easily
evaluated by Taylor-expanding the integrand
\be
S(\omega,q)= \frac{Av}{\pi}\frac{\omega^2-v^2q^2-9\d^2}{24\sqrt{3}\d^4}\ .
\label{smallx}
\ee
Eqn. \r{smallx} agrees with a result obtained for correlation
functions of the spin-1/2 ladder \cite{snt,GNT}.

Performing the remaining integral in \r{sfis} numerically, we obtain
the three magnon contribution to \r{sfis} as shown in
Fig.~\ref{fig:3is}. 
\begin{figure}[ht]
\noindent
\epsfxsize=0.45\textwidth
\epsfbox{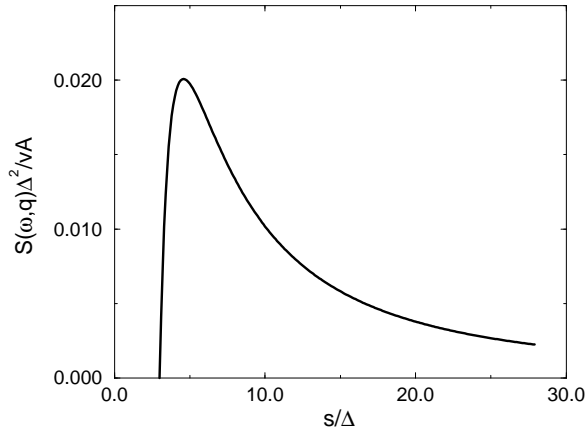}
\caption{\label{fig:3is}%
Result for the three magnon contribution to the dynamical
structure factor close to the antiferromagnetic point in Tsvelik's
Majorana fermion theory. Here $M$ is the magnon mass-gap, $v$ the spin
velocity and $s=\sqrt{v^2q^2+\d^2}$. 
} 
\end{figure}
The ratio of spectral weights of three-particle and one-particle
contributions to the structure factor \r{sfis} for frequencies
restricted to $\omega <20 M$ is roughly equal to
\be
\frac{I_3(\pi)}{I_1(\pi)}\approx 0.17\ .
\ee
Thus the three-particle scattering continuum is much stronger than in
the \nlsm!
The coupling between chains can be taken into account in complete
analogy with the \nlsm case.
Furthermore it is in principle possible to analytically study the
effects of the single-ion anisotropy in the Majorana fermion theory.

\section{Temperature}
\label{sec:temperature}

An additional complication in ${\rm CsNiCl_3}$ is that the ordering
temperature ${\rm 4.84K}$ is of the same order of
magnitude as the Haldane gap $\Delta\approx {\rm 15.4K}$. Clearly
temperature effects will therefore not be negligible in the temperature
range in which Neutron scattering experiments are conducted (5K-10K).
One way of taking them into account would be to use
recent results for finite-temperature correlation functions of the
Ising model \cite{ising3} in order to extend the analysis of section
\ref{sec:maj}. This is nontrivial, so that we constrain
ourselves to a rather preliminary discussion based on recent results
of \cite{sach}. There it was shown, that the finite temperature
lineshape of the one magnon peak at very low temperatures $T\ll \Delta$ is
given by 
\bea
&&S(\omega,q)= {\cal A}^\prime T\exp(-{\d}/{T})\nn
&&\quad\times\ 
\frac{1}{(\omega-\sqrt{\d^2+v^2q^2})^2+(1.2 T\exp(-\d/T))^2}\ .
\label{finiteT}
\eea
In other words, at temperatures much smaller than the Haldane gap the
lineshape is Lorentzian. Although \r{finiteT} does not really apply to
${\rm CsNiCl_3}$ because $T= {\cal O}(\Delta)$, it can be taken as an
indication that in the experimentally relevant temperature range, a
sizeable fraction of what used to be the one magnon spectral weight at
$T=0$ may get transferred to energies significantly above the single
magnon gap $\Delta$.

\section{Summary and Conclusions}
\label{sec:summary}
Using a low-energy effective field theory description (that needs to
be supplemented by numerical results in order to fix overall
normalisations) we have studied the dynamical structure factor for
weakly coupled integer spin Heisenberg chains. 
The three particle incoherent 
scattering continuum calculated in the O(3) \nlsm is found to be too
small to be observed in neutron scattering experiments. 

On the other hand, Tsvelik's Majorana fermion theory yields a very
strong three particle scattering continuum, which would be easily
observable in experiments. 
The main difference between the \nlsm and the Majorana fermion theory
is that in the latter the spin Hamiltonian contains strong biquadratic
interactions of the form
\be
\sum_n(\vec{S}_n\cdot\vec{S}_{n+1})^2\ .
\label{biquad}
\ee
The presence of such interactions in e.g. ${\rm CsNiCl_3}$ cannot a
priori be ruled out. As a matter of fact we believe that spin-1
materials generically have at least small interaction terms of the
form \r{biquad}. 

In the neutron scattering experiments \cite{mr} a significant incoherent
scattering continuum has been observed for several different
temperatures. These findings are incompatible with the \nlsm
results. However, the results of section \ref{sec:maj} indicate that
biquadratic exchange interactions lead to a transfer of spectral
weight from the coherent magnon peak to the incoherent scattering
continuum. One possible scenario for reconciling (field) theory with
the experimental findings is thus to postulate the existence of a
sizeable biquadratic exchange interaction in ${\rm CsNiCl_3}$. To the
best of our knowledge no detailed (beyond \cite{schmitt}) theoretical
study of the effects of biquadratic exchange interactions on the
dynamical structure factor has been carried out. We believe that it
would be interesting to do so.

As far as the present analysis is concerned, it has several
shortcomings that ought to be improved upon in future work. Firstly,
our analysis is restricted to zero temperature. As we argued above,
temperature effects are important in the experimentally realised range
of parameters. However, we believe it is unlikely that the finding of
\cite{mr} are due to temperature effects only.
Secondly, the analysis in the framework of the Majorana fermion theory
did not take into account the current-current interaction. The results
of section \ref{sec:maj} therefore should be regarded with some
caution. It would be very interesting to carry out a systematic
analysis of the effects of the current-current interaction.

Finally, we would like to mention that the analysis of section
\ref{sec:maj} applies with minor modifications to the case of the
spin-1/2 ladder as well. This is because the effective field theory
description is very similar \cite{snt}.

While this paper was being written, a preprint (cond-mat/9907431) by
M.D.P. Horton and I. Affleck appeared, in which the three magnon
contribution to the dynamical structure factor in the \nlsm was
calculated. Their results have a strong overlap with our section
\ref{sec:nlsm}.

\begin{center}
{\bf Acknowledgements}
\end{center}
I am grateful to Roger Cowley and Michel Kenzelmann
for many helpful discussions. This work was supported by the EPSRC
under grant AF/98/1081.

\end{narrowtext}

\end{document}